\documentclass[12pt]{article}

\usepackage{amsfonts,amsmath,amsxtra}

\newcommand{\beq}[1]{\begin{equation}\label{#1}}
\newcommand\eeq {\end{equation}}
\newcommand\bqa {\begin{eqnarray}}
\newcommand\eqa {\end{eqnarray}}
\newcommand\pr {\partial}

\newcommand{\eq}[1]{eq.(\ref{#1})\ }
\newcommand{\bear}{\begin{array}}
\newcommand{\enar}{\end{array}}

\begin{document}
\def\t{\theta}
\def\T{\Theta}
\def\w{\omega}
\def\ov{\overline}
\def\a{\alpha}
\def\b{\beta}
\def\g{\gamma}
\def\s{\sigma}
\def\l{\lambda}
\def\wt{\widetilde}

\hfill{AEI-2005-019}

\hfill{ITEP-TH-92-04}

\vspace{10mm}

\centerline{\bf \Large Simplicial vs. Continuum String Theory and
Loop Equations}

\vspace{10mm}

\centerline{{\bf Emil T.Akhmedov
}\footnote{email:{akhmedov@itep.ru}}}

\vspace{5mm}

\centerline{117218, Moscow, B.Cheremushkinskaya, 25, ITEP, Russia}

\centerline{and}

\centerline{Max--Planck--Institut f\"ur Gravitationsphysik,}
\centerline{Albert--Einstein--Institut, Am M\"uhlenberg 1,}
\centerline{14476 Golm, Germany}

\vspace{5mm}

\begin{abstract}
We derive loop equations in a scalar matrix field theory. We
discuss their solutions in terms of simplicial string theory
--- the theory describing embeddings of two--dimensional
simplicial complexes into the space--time of the matrix field
theory. This relation between the loop equations and the
simplicial string theory gives further arguments that favor one of
the statements of the paper hep-th/0407018. The statement is that
there is an equivalence between the partition function of the
simplicial string theory and the functional integral in a
continuum string theory --- the theory describing embeddings of
smooth two--dimensional world--sheets into the space--time of the
matrix field theory in question.
\end{abstract}

\vspace{10mm}

{\bf 1.} In this short note we give further arguments supporting
the observations made in \cite{Akhmedov:2004yb}. There we consider
matrix scalar field theory in the $D$--dimensional Euclidian
space:

\bqa Z = \int D \hat{\Phi}(x) \, D \hat{\bar{\Phi}}(x) \,
\exp\left\{- \int d^D x \, N \, {\rm Tr} \left[\frac12 \,
\left|\pr_\mu \hat{\Phi}\right|^2 + \frac{m^2}{2} \,
\left|\hat{\Phi}\right|^2 + \frac{\lambda}{3} \hat{\Phi}^3 + {\rm
c.c.}\right] \right\}, \label{qft}\eqa where $\mu = 1, \dots, D$,
$\hat{\Phi}$ is $N\times N$ matrix field in the adjoint
representation of $U(N)$ group: $\Phi^{ij}$, $i,j = 1, \dots, N$.
We choose this theory due to its simplicity (for our purposes) in
comparison with gauge and matrix theories with more involved
potentials. The problems of this theory due to the sign
indefiniteness of its potential are irrelevant for our
considerations: We consider this functional integral as a formal
series expansion over $\lambda$. All our considerations can be
easily generalized to the other matrix scalar and gauge theories.
In fact, one can always make a theory with cubic interactions out
of a theory with more involved interactions via insertions of
integrations over additional fields into the functional integral.

The functional integral (\ref{qft}) can be transformed into the
summation over the \underline{closed} two--dimensional simplicial
complexes\footnote{ Similar transformation has been done in
\cite{Freidel:2005bb} to establish a relation between the
no--gravity limit of the Ponzano--Regge theory and a
non-commutative field theory.} \cite{Akhmedov:2004yb}:

\bqa\log Z = \sum_{g=0}^{\infty} N^{\chi(g)} \sum_{V=0}^{\infty}
\lambda^V \sum_{{\rm graph; V,g \, fixed}}\, C'_{\rm graph}(V,g)
\times \nonumber \\ \times \left| \int_0^{+\infty} \prod_{n=1}^L
\frac{d\alpha_n}{\alpha_n^{g \, D}} \, e^{-\frac{\alpha_n \,
m^2}{2}} \, \int \prod_{a=1}^F d^D \vec{x}_a \, \exp\left\{-
\sum_{l=1}^L \frac{\alpha_l}{2} \,\left(\Delta_l \vec{x}\right)^2
\right\}\right|_{\rm graph}, \label{simplex}\eqa where $C'_{\rm
graph}(V,g)$ are some combinatoric constants defined in
\cite{Akhmedov:2004yb}; $F$ is the number of faces of the
\underline{fat} Feynman graph; $L$ is the number of links; $V$ is
the number of vertices and $g$ is the genus of the Feynman
diagram.

The summation in \eq{simplex} is taken over the graphs which are
dual to the Feynman diagrams \cite{Akhmedov:2004yb}. These graphs
represent triangulations of Riemann surfaces. In
\cite{Akhmedov:2004yb} we interpret the expression (\ref{simplex})
as the partition function of the closed simplicial string theory
--- the theory describing embeddings of two--dimensional
simplicial complexes into the space--time of the matrix field
theory. In the context $\alpha$'s are related to the components of
the two--dimensional metric \cite{Akhmedov:2004yb}.

 Furthermore, in \cite{Akhmedov:2004yb} we argue that there is no need to take a
continuum limit in \eq{simplex}: There should be a continuum
string theory\footnote{I.e. the theory describing embeddings of
the smooth two--dimensional world--sheets into the space--time of
the matrix field theory in question.} whose functional integral is
equal to \eq{simplex}. In this paper we give further arguments
supporting this idea. We propose equations which are solved via
the simplicial \underline{open} string theory ``functional
integral''. On the other hand these equations have a natural
interpretation as constraint equations in a two--dimensional field
theory containing gravity.

To present the idea of our argument, let us consider the case of
the relativistic particle. The path integral for the latter solves
the following equation \cite{Polyakov:ez}:

\bqa\left(-\Delta + m^2\right)\, G(\vec{x}, \, \vec{x}') =
\delta(\vec{x} - \vec{x}').\label{first}\eqa One can also obtain a
simplicial integral solution to this equation
\cite{Akhmedov:2004yb} as follows. The solution of \eq{first} can
be represented as:

\bqa G(\vec{x}, \, \vec{x}') = \int d^D \vec{p} \,\, e^{{\rm i} \,
\vec{p} \, (\vec{x} - \vec{x}')} \, \frac{1}{\vec{p}^2 + m^2} =
\frac{1}{\Lambda}\int d^D \vec{p} \,\, e^{{\rm i} \, \vec{p} \,
(\vec{x} - \vec{x}')} \, e^{-\log{\frac{\vec{p}^2 +
m^2}{\Lambda}}} = \nonumber \\ =
\frac{1}{\Lambda}\sum_{L=0}^{\infty} \frac{(-1)^{L}}{L!} \,  \int
d^D \vec{p} \,\, e^{{\rm i} \, \vec{p} \, (\vec{x} - \vec{x}')}
\,\left[\log{\frac{\vec{p}^2 + m^2}{\Lambda}}\right]^L = \nonumber
\\ = \frac{1}{\Lambda}\sum_{L=0}^{\infty} \frac{(-1)^{L}}{L!} \,
\int d^D \vec{p} \,\, e^{{\rm i} \, \vec{p} \, (\vec{x} -
\vec{x}')} \, \prod_{l=1}^L \, \int_0^{+\infty} \, \frac{d
e_l}{e_l} \left(e^{-\frac{\vec{p}^2 + m^2}{2} \, e_l} -
e^{-\frac{\Lambda}{2} \, e_l} \right), \label{leading}\eqa where
$\Lambda$ is the cutoff. If we drop all terms containing
$\exp\left\{- \Lambda \, e/2\right\}$ in \eq{leading}, we obtain
the divergent expression which can be represented in the form
\cite{Akhmedov:2004yb}:

\bqa G_{\rm div}(\vec{x}, \, \vec{x}') \propto \sum_{L=0}^{\infty}
\frac{(-1)^{L}\, C_L}{L!} \, \int_0^{+\infty} \prod_{n=1}^{L}
\frac{d e_n}{e^{D/2+1}_n} \int \prod_{i=1}^L d^D \vec{y}_i \,
\exp\left\{- \frac12 \, \sum_{l=0}^{L} \left[\frac{\left(\Delta_l
\vec{y}\right)^2}{e_l} + m^2 \, e_l \right]\right\}.
\label{main}\eqa In each member of the sum here $\vec{y}_0 =
\vec{x}$, $\vec{y}_{L+1} = \vec{x}'$ and $C_L$ are easily
computable constants dependent on $L$.

The formula (\ref{main})  contains the summation over the
one--dimensional geometries. In fact, it contains the summation
over all discretizations/triangulations ($L$) of the
world--trajectory and the integration over all
\underline{one--dimensional} distances ($e$'s) between the
vertices ($y$'s). The summation over the embeddings of the
simplicial complexes is presented by the summation over the number
of vertices ($L$) and the integration over all their possible
positions, i.e. over
--- $y$'s.

Thus, \eq{main} is, so to say, a simplicial particle theory ``path
integral'' which formally solves \eq{first}, but demands a
regularization. At the same time \eq{leading} suggests a natural
regularization of the simplicial partition function (\ref{main})
and rigorously relates it to the differential \eq{first}.
Moreover, as we see one does not have to take a continuum limit in
\eq{main} and this ``simplicial path integral'' after the
regularization is equivalent to the regularized standard path
integral for the relativistic particle.

Below we argue that the same story should happen in the case of
simplicial string theory and, possibly, for the higher dimensional
simplicial brane theories. To obtain a proper theory of the latter
kind one should both sum over the (multi--dimensional)
triangulations and integrate over the sizes of the links: this
gives the summation over all internal geometries, which in usual
functional integrals is represented by the integration over all
metrics divided by the volume of the group of diffeomorphisms.

{\bf 2.} Once the relation between \eq{qft} and \eq{simplex} is
established, one of the natural generalizations of \eq{first} to
two--dimensions can be represented by the loop equations
\cite{Makeenko:1980vm} in the matrix field theory. In this section
we derive the loop equations for the theory (\ref{qft}) and
discuss their obvious solution in terms of the simplicial open
string theory. Such a string theory follows from the expansion in
Feynman diagrams of an analog of the Wilson's loop correlation
function \cite{Akhmedov:2004yb}. As we will see, these loop
equations have a natural interpretation as constrained equations
on the functional integral for a continuum string theory.
Obviously the latter should be equivalent to the simplicial string
theory partition function in the same way as it happens in the
case of the relativistic particle.

 Thus, we would like to consider Ward type identities for the correlation function of
the Wilson loop operator:

\bqa W(C) = {\rm Tr P} \exp\left\{ - \oint_C ds \,
\sqrt{\dot{\vec{x}}^2(s)} \, \hat{\Phi}\left[x(s)\right]\right\},
\eqa where $C$ is a loop in the space--time, which is represented
by the map $x(s)$. However, one can obtain closed\footnote{Means
equations which include no other kinds of operators except the
loop ones.} loop equations for such an operator only in the theory
with the Lagrangian \cite{Makeenko:1988hm}:

\bqa L = \frac12 \, {\rm Tr} \left|\pr_\mu \hat{\Phi}\right|^2\eqa
or with the Lagrangians following from the reduction of the
Yang--Mills theory. To obtain closed loop equations for the theory
(\ref{qft}) we suggest to consider the loop operator as follows:

\bqa W(C,e) = {\rm Tr P} \exp\left\{ - \oint_C ds \, e(s) \,
\hat{\Phi}\left[x(s)\right]\right\}. \eqa As well there is the
operator $\bar{W}(C,e)$ which depends on $\bar{\Phi}$ and the same
$e$ --- real--valued square root of the one--dimensional internal
metric on the interval of $s$.

Let us define the loop space Laplace operator as in
\cite{Polyakov:ez}:

\bqa\frac{\pr^2}{\pr x^2(s)} = \int_{s-0}^{s+0} ds'
\frac{\delta^2}{\delta x_\mu(s)\, \delta x_\mu(s')}.\eqa Then it
is straightforward to see that \cite{Polyakov:ez},
\cite{Makeenko:1980vm}:

\bqa\left(- \frac{\pr^2}{\pr x^2(s)} + m^2 \, e(s) \,
\frac{\pr}{\pr e(s)}\right) W(C,e) + \lambda \, e(s) \,
\frac{\pr^2}{\pr e^2(s)} \bar{W}(C,e) = \nonumber \\ = e(s) \,
{\rm Tr P} \, \left\{\left(-\pr_\mu^2\hat{\Phi} + m^2 \,
\hat{\Phi} + \lambda \, \hat{\bar{\Phi}}^2\right)\, \exp\left\{ -
\oint_C ds \, e(s) \,
\hat{\Phi}\left[x(s)\right]\right\}\right\}.\eqa Similarly one has
the complex conjugate equation. To find the RHS of this expression
(after the averaging over all field configurations), let us
consider the equality\footnote{Here $\hat{\Phi} = \Phi^a \, T^a$
and $T^a$, $a=1, ..., N^2$ are the generators of $U(N)$.}:

\bqa 0 = \int D \hat{\Phi}(x) \, D \hat{\bar{\Phi}}(x) \,
\frac{\delta}{\delta \bar{\Phi}^a} \,\left( \exp\left\{- \int d^D
x \, N \, {\rm Tr} \left[\frac12 \, \left|\pr_\mu
\hat{\Phi}\right|^2 + \frac{m^2}{2} \, \left|\hat{\Phi}\right|^2 +
\frac{\lambda}{3}
\hat{\Phi}^3 + {\rm c.c.}\right] \right\} \right.\times \nonumber \\
\times \left. {\rm Tr P} \exp\left\{ - \oint_C ds \, e(s) \,
\hat{\bar{\Phi}}\left[x(s)\right]\right\}\right).\eqa From this we
obtain:

\bqa \left\langle \left(-\pr_y^2 \Phi^a(y) + m^2 \,\Phi^a(y) +
\lambda \, \left[\bar{\Phi}^2\right]^a(y)\right)\, \exp\left\{ -
\oint_C ds \, e(s) \, \hat{\Phi}\left[x(s)\right]\right\}
\right\rangle = \nonumber
\\ = - \left\langle \oint ds \, e(s) \, \delta\left[y - x(s)
\right] \, {\rm P} \, \exp\left\{ - \int_x^y dt \, e(t) \,
\hat{\bar{\Phi}}\left[x(t)\right]\right\} T^a \exp\left\{ -
\int_y^x dt \, e(t) \, \hat{\bar{\Phi}}\left[x(t)\right]\right\}
\right\rangle. \eqa Here the LHS appears from the variation over
$\bar{\Phi}^a$ of the exponent of the action and the RHS appears
from the variation of $\bar{W}(C,e)$.

Hence, we obtain:

\bqa\left(- \frac{1}{e(s)} \, \frac{\pr^2}{\pr x^2(s)} + m^2 \,
\frac{\pr}{\pr e(s)}\right) \left\langle W(C,e)\right\rangle +
\lambda \, \frac{\pr^2}{\pr e^2(s)} \left\langle
\bar{W}(C,e)\right\rangle= \nonumber \\ = \oint ds' \, e(s')
\,\delta\left[x(s) - x(s') \right] \, \left\langle
\bar{W}(C_{xx'},\,e) \, \bar{W}(C_{x'x}, \,
e)\right\rangle\label{loopeq}\eqa and the complex conjugate
equation. In \eq{loopeq} we use:

\bqa \sum_a T^a_{ij} \, T^a_{mn} = \delta_{in} \, \delta_{jm} \eqa
and $C^{xx}_{x'x'} = C_{x'x} \cup C_{x x'}$. The RHS of
\eq{loopeq} does not vanish if the contour $C^{xx}_{x'x'}$ (which
is just $C$ with two designated points $x=x(s)$ and $x'=x(s')$)
has self--intersection at $x = x'$ \cite{Polyakov:ez},
\cite{Makeenko:1980vm}.

 The solution of \eq{loopeq} via the
expansion in powers of $\lambda$ of the correlation function
$\langle W(C,e) \rangle$ looks as follows:

\bqa \log \left\langle W(C,e) \right\rangle = \nonumber
\\ = \sum_{E=2}^{\infty} \int_0^{2\pi} ds_1 \, e(s_1) \dots
\int_0^{s_{E-1}} ds_{E} \, e(s_{E}) \, \sum_{g=0}^{\infty}
N^{\chi(g)} \sum_{V=0}^{\infty} \lambda^V \,\sum_{\rm graph; V, g,
E \, fixed} C_{\rm graph}(E,V,g)
\times \nonumber \\
\times \left| \int_0^{+\infty} \prod_{n=1}^L d\alpha_n \, \int
\prod_{i=1}^V d^D \vec{y}_i \, \int \prod_{m=1}^L d^D \vec{p}_m \,
\exp\left\{- \sum_{l=1}^L\left[\frac{\alpha_l \, \left(\vec{p}_l^2
+ m^2\right)}{2} - {\rm i}\, \vec{p}_l \left(\Delta_l
\vec{y}\right) \right]\right\}\right|_{\rm
graph},\label{graph}\eqa where $C_{\rm graph}(E,V,g)$ are some
combinatoric constants and in the exponent on the RHS among the
$y$'s there are $y(s_1) , \dots, y(s_E)$ over which the
integration is not taken and they are sitting on the contour $C$.
The first sum on the RHS is taken over their number. The summation
over ``graph'' in \eq{graph} means the summation over the Feynman
diagram contributions to the correlation function in question.
Accordingly, $V$ is the number of interaction vertices; $L$ is the
number of propagators; $y$'s are positions of the vertices; $p$'s
are momenta running over the propagators; $\alpha$'s are Schwinger
parameters and $g$ is the genus of the fat Feynman diagram.

 Performing the transformation of \cite{Akhmedov:2004yb}, we obtain:

\bqa \log \left\langle W(C,e) \right\rangle = \sum_{E=2}^{\infty}
\int_0^{2\pi} ds_1 \, e(s_1) \dots \int_0^{s_{E-1}} ds_{E} \,
e(s_{E}) \, \sum_{g=0}^{\infty} N^{\chi(g)} \sum_{V=0}^{\infty}
\lambda^V \times \nonumber
\\ \times \sum_{\rm graph; V,g, E \, fixed} C'_{\rm graph}(E,V,g) \,
\left|\int_0^{+\infty} \prod_{n=1}^L
\frac{d\alpha_n}{\alpha_n^{(2g+1)\, D/2}} \, e^{-\frac{m^2 \,
\alpha_n}{2}}\,  \int \prod_{a=1}^F d^D \vec{x}_a \, \exp\left\{-
\sum_{l=1}^L \frac{\alpha_l}{2} \,\left(\Delta_l \vec{x}\right)^2
- \right. \right.  \nonumber \\
\left. \left. - \sum_{f=1}^E
\frac{\vec{y}^2(s_f)}{2}\,\sum_{s,s'=1}^{2g+1} \omega_f^{(s)} \,
\frac{1}{\sum_{l=1}^L \alpha_l \, \omega_l^{(s)}\,
\omega_l^{(s')}} \, \omega_f^{(s')} + {\rm i} \, \sum_{f=1}^E
\Delta_f\vec{x} \, \vec{y}(s_f)\right\}\right|_{\rm graph}.
\label{main1}\eqa Here $\omega_l^{(s)}$, $s = 1, ..., 2g+1$ are
the values on the $l$-th link of the closed (but not exact)
one--forms on the genus $g$ simplicial complex with one boundary.
These simplicial complexes are defined by the dual graphs to the
Feynman diagrams: Now the sum in \eq{main1} is taken over these
dual graphs rather than the Feynman diagrams themselves. $C'_{\rm
graph}$ is different from $C_{\rm graph}$ by a factor of the
determinant of some matrix \cite{Akhmedov:2004yb}.

  The main difference between \eq{loopeq} and \eq{first} is that
the former one is the non-linear equation. But dropping the RHS of
\eq{loopeq} (and putting the functional $\delta$--function
instead), we obtain the standard linear Wheeler--DeWitt equation
in a two--dimensional gravity theory coupled to the matter fields
($x$). Both loop and Wheeler--DeWitt equations are not well
defined due to their divergences \cite{Polyakov:ez}. As the
result, the solution of such equations in terms of
two--dimensional functional integral is not known.

Note that the UV divergences of the quantum field theory in
\eq{qft} acquire a clear interpretation in the simplicial string
theory description (\ref{main1}). These divergences are just due
to the boundaries in the space of all metrics, i.e. when some of
the $\alpha$'s vanish, which corresponds to the situations in
which some of the triangles in the dual graph to the Feynman
diagram degenerate into links \cite{Akhmedov:2004yb}. The natural
regularization of \eq{main1} is analogous to the one presented in
\eq{leading} for the case of particle. It is nothing but the
regularization which follows from the insertion of the integration
over the ghost Pauli-Villars fields into the functional integral
of the matrix field theory. The addition of these fields sets an
obvious regularization of the loop equations, but one needs a
renormalized version of these equations rather than just their
regularization \cite{Polyakov:ez}. This is the subject for another
work (see \cite{Akhmedov:1998vf} for the attempts of understanding
this point).

{\bf 3.} We have considered nonstandard loop variables in the
scalar matrix field theory. These loop variables depend on both
loops in the target space and internal one--dimensional metrics
and obey loop equations. The equations represent a non--linear
generalization of the Wheeler--DeWitt equations in a
two--dimensional gravity theory interacting with matter. There is
an obvious solution to these equations in terms of the partition
function of an open simplicial string theory. We argue that there
should be a continuum string theory solution to the same equations
which is exactly equivalent to the simplicial one. The only
obstacle which can appear in formulating such a continuum string
theory is that for generic values of $\lambda$ it can happen that
its functional integral will contain an integration measure for
the metrics which does not follow from a local norm.

I would like to acknowledge valuable discussions with Yu.Makeenko,
M.Zubkov, F.Gubarev, V.Shevchenko, N.Amburg, T.Pilling and
V.Dolotin. I would like to thank H.Nicolai, S.Theisen,
A.Kleinschmidt, M.Zamaklar and K.Peeters for the hospitality
during my visit to the MPI, Golm. This work was done under the
partial support of grants RFBR 04--02--16880, INTAS 03--51--5460
and the Grant from the President of Russian Federation
MK--2097.2004.2.

\end{document}